\documentclass[aps,twocolumn]{revtex4}

\usepackage{graphicx}
\usepackage{rotating}
\newcommand{\be}{\begin{equation}}
\newcommand{\ee}{\end{equation}}

\input{epsf}

\begin{document}
\title{Dynamic equilibrium Mechanism for  Surface Nanobubble Stabilization}
\author{Michael P.\ Brenner$^1$ and Detlef Lohse$^2$}
\affiliation{$^1$ School of Engineering and Applied Sciences,
Harvard University, Cambridge, MA 02138, USA\\
$^2$
Physics of Fluids Group,
Faculty of Science and Technology, MESA+ and Impact
Institutes, University of Twente, 7500 AE Enschede, The
Netherlands
}

\begin{abstract}
Recent experiments have convincingly demonstrated the existence of
surface nanobubbles on submerged hydrophobic surfaces. However, classical theory dictates that small gaseous bubbles quickly dissolve because
their large Laplace pressure causes a diffusive outflux of gas.  Here we suggest that the bubbles are stabilized by a continuous influx of gas near the contact line,
due to the gas attraction towards hydrophobic walls
(Dammer \& Lohse, PRL96, 206101 (2006); 
Zhang {\it et al.}, PRL98, 136101 (2007);
Mezger {\it et al.}, J.\ Chem.\ Phys.\ 128, 244705 (2008)). 
This influx balances the outflux and allows for a meta-stable
equilibrium, which however vanishes in thermodynamic 
equilibrium. Our theory predicts
the equilibrium radius of the surface nanobubbles,
as well as the  threshold for surface nanobubble
formation as a function of hydrophobicity and gas concentration.
\end{abstract}

\date{\today}

\maketitle

Various recent 
studies 
 have  revealed the existence of
nanoscopic soft domains at the liquid-solid interface
\cite{bal03,vin95,tyr01,att03,hol03,sim04,zha04,agr05,net05,zha06,zha06b,zha07b,yan07,bor07,yan08,zha08}.
As atomic force microscopy (AFM) shows, 
these
soft domains resemble spherical caps with heights of  $\approx 10\,{\rm nm}$ and diameters of $ 2 R \approx 100\,{\rm nm}$, corresponding to a small
contact angle
(on the gas side) of $\theta  \approx 10^\circ$,
see figure 1 for a sketch.
The most consistent interpretation \cite{tyr02} of these soft domains
 is that they
are {\em
  surface nanobubbles}, i.e., nanoscale gas bubbles located at the
liquid-solid interface.
Spectroscopic studies \cite{zha07,zha08} 
and X-ray reflectivity measurements \cite{mez06,mez08,poy06}
experimentally confirm 
the presence of gas at the wall.
Moreover, 
the size and density of these objects depend on the  dissolved gas
concentration and they
disappear completely when the liquid is degassed 
\cite{zha06b,zha06,zha07,swi04,yan07}. 
CO$_2$ bubbles have a much shorter lifetime (only 1-2 hours) as compared 
to air
bubbles, due to the much lower pressure of CO$_2$ in the atmosphere
and its better solubility in water \cite{zha08}. 
  The existence of these surface nanobubbles has potentially great technological importance, as they have been shown to result in fluid slip
in hydrophobic surfaces, leading to a potentially large reduction in fluid dissipation for flows in small devices \cite{lau05,fan05,vin06}.

Observations of long living surface nanobubbles flatly contradicts the classical theory of bubble stability \cite{eps50}. Small bubbles  have large internal gas pressures, in order to balance the compressive action of surface tension. For a bubble in water of radius  $R=50$nm,
which with above  contact angle $\theta = 10^\circ$ corresponds
to a radius of curvature $R_c = R/\sin\theta \approx 250$nm, and 
surface tension $\gamma=73$mN/m ($20^\circ$ temperature), the gas pressure 
according to classical macroscopic theory 
(neglecting the disjoining pressure \cite{gen03})
is  $P_{gas} = 2\gamma/R_c
= 2\gamma \sin\theta/R=0.58$MPa.
In employing Laplace's equation 
we also neglect the bubble-stabilizing contribution from hydroxide ion 
adsorption on the bubble surface \cite{kar02}. If the resulting charge 
density is constant, this ion adsorption contributes a 
 radius-independent term to $P_{gas}$, which 
is neglibible for small enough bubbles.
Henry's law then dictates that the concentration of gas at the bubble surface is $c(R)=c_0 P_{gas}/P_0 $, where $c_0$ is the saturated concentration of gas at atmospheric pressure $P_0$. This is necessarily much larger than the gas concentration $c_\infty\le c_0$ far away, leading to a diffusive outflux of gas and bubble dissolution \cite{eps50,lju97}. {\it If} the contact angle $\theta$ remains finite for $R\to 0$ and assuming that 
 Laplace's law remains valid at molecular scales, then
$c(R)$ formally diverges, signalling the breakdown
of the continuum approach.

Various suggestions have been made to
explain the long lifetime of surface nanobubbles. Among these, 
the reduction of surface
tension for large curvature  on small scales 
\cite{alb00,mor03} 
(operative for bubbles smaller than $\sim 5$nm), 
the oversaturation of the liquid 
 around the nanobubbles with gas \cite{zha04,zha08}, the effect of 
induced charges in the Debye layer around the bubble interface
 \cite{tse08}, or the stabilization of bubbles due to contaminant 
molecules in the water \cite{dre08}, similarly as has been suggested
for microbubbles in contaminated water \cite{joh81}.
 In this paper we propose an alternative mechanism for
 bubble stabilization, namely, that the gas outflux is compensated by 
 gas influx at the contact line.

The diffusive mass outflux from a surface nanobubble   is given by
\be
J_{out}(R) = D \int \ dA {\bf n}\cdot \nabla c \approx 2\pi D\int_0^R \ r dr \partial_zc,
\label{jout}
\ee
where
$D$ is the diffusion constant of gas in the liquid (typically 
$10^{-9}$m$^2$/s)
 and ${\bf n}$ is the normal vector to the droplet surface. 
We  have assumed that the bubble is nearly flat, so that the 
diffusive gradient is primarily in the $\hat{z}$ direction, 
perpendicular to the solid surface.
This diffusive flux can then be evaluated 
 by solving the steady state diffusion equation in the liquid. The result is
\be
J_{out} (R) = \pi R D  (c(R)- c_\infty )
= \pi R D \left( {c_0\over P_0} \,{2\gamma \sin\theta \over R} - c_\infty
\right).
\label{jout2}
\ee
The volume flux rate 
$j_{out}(R)=J_{out}(R)/c(R)$ is then given by
\be
j_{out}(R) = {J_{out}(R)\over c(R)}
= \pi R D \left(
1-{c_\infty \over c(R)}
\right).
\label{jout3}
\ee
The volume of the bubble thus decreases linearly with time, typically
with a volume flux rate
$\approx \pi RD \approx \pi \cdot \ 50\cdot 10^{-9}\ \cdot 10^{-9}$m$^3$/s,
when assuming complete degasing. 
For an initial bubble of radius $50$nm, this implies a dissolution timescale of $\sim 1\mu$s.

Stabilization against dissolution
 requires a physical effect to cancel this diffusive outflux. Although intermolecular forces might modify the liquid surface tension from its 
macroscopic value \cite{gen03,mor03},  the energetic cost of creating surface energy ensures that the gas pressure in the bubble is always higher than that of the surrounding liquid, and hence  diffusive outflux  necessarily persists even when interactions with the solid surface are accounted for.
Stabilization can be achieved however, by mechanisms causing an {\it influx}
 of gas into the bubble. Such an influx need not occur uniformly across the bubble surface, but can be spatially concentrated.
In particular,
it is quite natural to consider  influx mechanisms  near the contact line, where
intermolecular forces are most significant. For a sketch of the gas flow
directions, we refer to figure 1.
The magnitude of any contact line dominated influx increases with the circumference of the bubble, and hence could compete effectively with the diffusion mediated outflux eqs.\ (\ref{jout2}) and (\ref{jout3}).

What is the origin of such an influx?
Recent MD simulations \cite{dam06,luz05,bra08}
of gas dissolved in water in contact with a surface
demonstrated that on hydrophobic surfaces there is gas 
enrichment near solid walls. 
Such an 
enhanced gas concentration at hydrophobic walls has 
been confirmed 
spectroscopically \cite{zha07b,zha08} 
and with X-ray reflectivity measurements \cite{mez06,mez08,poy06};
also neutron reflectivity experiments \cite{ste03} suggest it.  
In the MD simulations of ref.\ \cite{dam06}, for sufficiently hydrophobic walls the concentration near walls can exceed more than two orders of magnitude above the concentration in bulk liquid.  Physically, this enrichment occurs because there is  a potential $\phi(z)$ attracting solute molecules to the wall. The equilibrium concentration of the solute is thus determined by the balance between diffusion and attraction, according to
$D \frac{dc}{dz}= -\frac{1}{\zeta}\frac{d\phi}{dz} c$
where  $\zeta$ is the mobility of the solute. The Einstein relation
$D=k_B T/\zeta$ then implies the equilibrium distribution of solute as $c(z)=c_0 \exp ({-\phi(z)/k_B T}).$
A hundredfold concentration of solute molecules near the wall
as found for the case  studied in ref.\ \cite{dam06}
implies that  $\phi(z)$ decreases by about $4 k_B T$ near the wall, over a molecularly determined length scale. In general, we assume that the energy gain $\Delta\phi$
of a solute molecule at the wall
equals
$\Delta \phi = s k_BT$.

\begin{figure}
\begin{center}
\includegraphics*[width=6cm,angle=-90.]{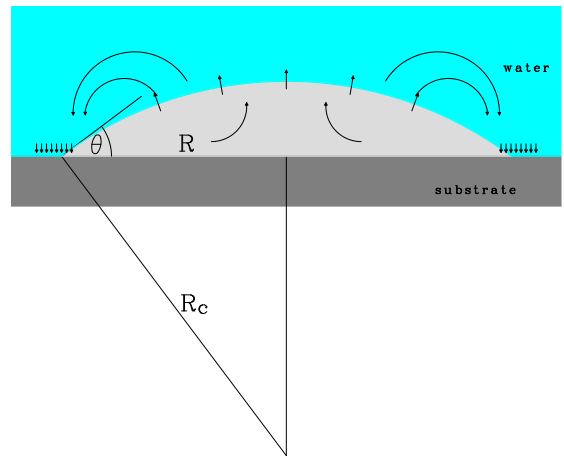}
\caption{
Sketch of gas outflux and influx into a surface nanobubble and
definition of the contact angle $\theta$,  the surface nanobubble
radius $R$, and the radius of curvature $R_c$.
}
\label{sketch}
\end{center}
\end{figure}

\begin{figure}
\begin{center}
\includegraphics*[width=8cm]{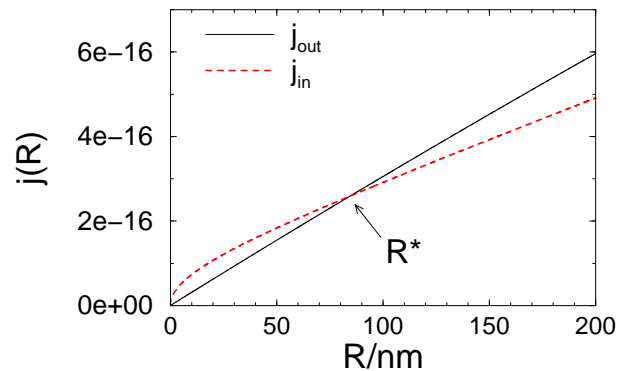}
\caption{
Gas outflux $j_{out}$ (black-solid, from eq.\ (\ref{jout3}))
and influx $j_{in}$ (red-dashed, from eq.\ (\ref{result3}))
into the surface
nanobubble as function of bubble radius $R$. The
crossing point defines the equilibrium radius $R^*$.
If the slope of $j_{out}$ at $R=0$ is larger than that
of $j_{in}$, no surface nanobubbles can emerge.
For this illustrative plot we used
eq.\ (\ref{line-tension2}) with 
$\delta = 70$nm, $\theta_\infty = 50^\circ$, and $\theta_0 = 0^\circ$, 
and the values
$s=0.36$ 
for the relative strength of the attraction potential,
$c_\infty/c_0 = 0.25$ for the relative gas concentration, and
$D=10^{-9}m^2/s$ (which only scales the y-axis).
For these data the stable equilibrium radius is $R^*=85$nm.
}
\label{fluxes}
\end{center}
\end{figure}

\begin{figure}
\begin{center}
\vspace*{0.3cm}
\includegraphics*[width=7.5cm]{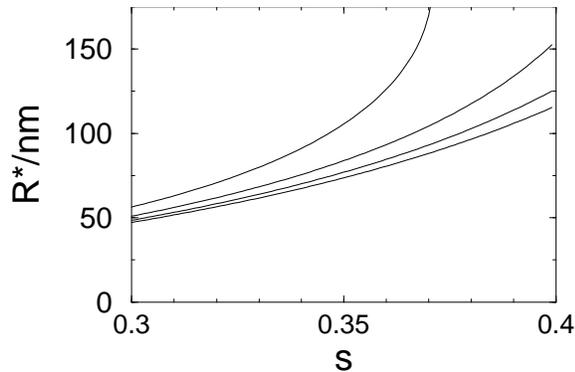}
\caption{
Dependence of equilibrium radius $R^*$ on the wall attraction strength
$s$ for relative gas concentrations
$c_\infty/c_0=$ 0.1, 0.25, 0.5, 0.95, bottom to top. 
}
\label{r-star-of-various}
\end{center}
\end{figure}

 Solute molecules at a gas-liquid interface are attracted to the wall and hence driven inside the bubble, see again figure 1.
The size of the gas influx  $J_{in}$ is given by
\be
J_{in}(R) \approx  2\pi \int_0^R r \ dr \ c(r)\frac{1}{\zeta}{\frac{d\phi(z=h(r))}{dz}},
\label{influx}
\ee
Here we have again approximated the bubble shape $h(r)$ as nearly flat. Since
the attractive force drops off quickly with distance, the flux is only appreciable near the contact line. Using the fact that near the contact line 
$h(r)=(R-r) \tan\theta$
we can approximate equation (\ref{influx}) to obtain the mass influx
$
J_{in}(R) 
\approx 2\pi R c(R)
\Delta \phi/
(\zeta \tan\theta ), 
$
or the volume influx
\be
j_{in} (R) = {J_{in}(R) \over c(R)}
\approx \frac{2\pi s D R }{ \tan\theta (R)}.
\label{result3}
\ee

As anticipated above, eq.\ (\ref{result3}) suggests that the influx scales linearly with $R$, exactly the same scaling as the diffusive outflux.
However, recent measurements \cite{che03,che06,jak04,li96},
including some on surface nanobubbles \cite{kam08a,kam08b}, 
have shown that for sufficiently small bubbles or droplets, the contact angle $\theta$ depends on the contact line curvature and thus on the bubble size. 
This can be described through
a modified Young-Dupr\'e equation
\be
\cos(\pi-\theta)=\cos(\pi-\theta_\infty) - \frac{C}{R+\delta},
\label{line-tension}
\ee
where  $C$ and $\delta$ are constants with the dimensions of length, and $\theta_\infty$ is the usual macroscopic equilibrium contact angle. A popular (though disputed \cite{che03}) model assumes that the correction arises from  a line tension $\tau$ of the contact line \cite{gay87}, in which case $C=\tau/\gamma$,where $\gamma$ is the liquid surface tension.
The ratio $C/ \delta$ is set by the contact angle $\theta_0$
for $R\to 0$, namely $C/ \delta = \cos (\pi - \theta_\infty)
-\cos (\pi - \theta_0 )$, so that one can rewrite (\ref{line-tension}) as
\be
\cos\theta=\cos\theta_\infty
- \frac{\cos \theta_\infty -\cos \theta_0 }{1+R/\delta}.
\label{line-tension2}
\ee
The length scale $\delta$ is the onset-scale of microscopic corrections
to the macroscopic contact angle. 
The exact form of eq.\ (\ref{line-tension2}) 
and the exact values of the parameters
are not relevant
in the context of this paper; all what is needed is that $\theta(R)$ 
decreases with decreasing bubble radius $R$,
which indeed is observed in refs.\ \cite{che03,che06,kam08a,kam08b}.
We take $\delta =70$nm, 
$\theta_\infty = 50^\circ$, and
$\theta_0 = 0^\circ$. 
Note that the behavior $\theta (R\to 0) =0$ 
leads to a stabilization
of small surface nanobubbles, as their curvature becomes small.
In the limiting case $R\to 0$ it even vanishes and the numerically found
\cite{dam06} case of a mono- or bilayer of gas molecules between the 
surface and the liquid is recovered.

The outflux
$j_{out}(R)$ and the influx
$j_{in}(R)$ are shown in figure 2 for some typical
parameters. In that case, a {\it stable} dynamic equilibrium radius $R^*$
defined through $j_{out}(R^*) = j_{in}(R^*)$ exists.
Note that indeed various studies have revealed
the existence of a {\it prefered} radius of 
the surface nanobubbles \cite{sim04,zha08}, 
which
depends on the gas concentration \cite{zha08}.
For bubbles smaller than the equilibrium radius,
 the influx overcompensates the outflux,
for larger bubble the outflux wins. If the gas concentration $c_\infty$
decreases, the outflux increases and the equilibrium radius is becoming
smaller, in agreement with experimental observations on nanobubbles
\cite{yan07}. If, on the other hand, the surface gets more hydrophobic
and $\theta_\infty$  decrease and/or the attraction potential
$s$ increases, the equilibrium radius
is shifted towards larger values, again in agreement with experimental
observations.

The necessary condition for
a stable dynamic equilibrium and therefore for stable nanobubbles
to exist is that at small bubble sizes $R\to 0$
the influx $j_{in}(R)$ is larger  than the outflux $j_{out}(R)$.
This implies the condition
\be
s  > {1\over 2} \tan \theta_0
\left(1-\frac{c_\infty}{c(R\to 0)} \right) \approx
{1\over 2} \tan \theta_0 
\label{condi2}
\ee
for small gas concentration $c_\infty \ll c(R\to 0)$.
The condition (\ref{condi2}) is satisfied for  sufficiently hydrophobic surfaces, where $\tan(\theta(R\to 0))$ is small enough and when the solute is attracted to the wall, so that $s$ is large enough.
If condition (\ref{condi2}) is fulfilled, there also is an
unstable equilibrium at $R^*=0$, allowing for spontaneous nanobubble formation.
If that condition is not fulfilled, the equilibrium at $R^*=0$ becomes stable
and no surface nanobubbles form, in spite of an enhanced gas concentration
very close to the surface. These may have been the conditions of the
X-ray reflectivity experiments of ref.\ \cite{mez06}, where a gas 
accumulation, but no nanobubbles, 
have been seen. 

How does the prefered bubble radius $R^*$ depend on the material 
properties and the control parameters? 
To illustrate this dependence, we show
 $R^*(s,c_\infty/c_0)$
resulting from  our simple model 
(eqs.\ (\ref{jout3}, (\ref{result3}), and (\ref{line-tension2}), 
see fig.\ \ref{r-star-of-various}. 
The dependence of $R^*$ 
on the attraction strength $s$ 
is as expected rather strong -- slight (chemical or structural)
inhomogeneities in 
the surface will therefore result into some  distribution
 in the prefered 
radius, as indeed experimentally seen in refs.\ \cite{sim04,zha08}.
In contrast, away from saturation, 
the dependence on the exact value of the relative gas concentration
$c_\infty/c_0$
is relatively weak.

Some comments on the driving mechanism:
While the gas diffuses in the liquid, there is a gas flow from the
contact line towards the surface inside the
bubble. 
To avoid conflict with 
the second 
law of thermodynamics, 
such a state can only be transient.
We do not know the origin of the non-equilibrium; it 
could be caused by temperature gradients, chemical effects,
 or  local oversaturation.
In principle the observation method itself
(e.g.\ AFM or optical 
detection of the nanobubbles) could be the origin
of the equilibrium distortion.
If the system were isolated, the second law of thermodynamics
requires that in the long term (possibly hours or even days) the
driving mechanism would expire and the surface nanobubbles therefore dissolve,
i.e., nanobubbles would be a transient phenomenon.
Unfortunately, long-term observations of surface nanobubbles
in closed sytems
have to our knowledge not yet been carried out.
Note that the required non-equilibrium situation must 
also reflect in the
chemical potential $\mu$, as in equilibrium 
gas flow can only occur along a gradient in 
$\mu$, and a circular flow would not be possible.

A method to create  {\it controlled} non-equilibrium conditions is
to generate surface nanobubbles through electrolysis \cite{zha06b,yan08b}.
In ref.\ \cite{yan08b} it has been shown that 
several tens of seconds after switching on
the potential, these surface nanobubbles do not further grow, in spite of a 
nonzero current.
This observation suggests that the nanobubbles indeed have 
achieved a {\it dynamic} stable
 equilibrium, with the Laplace pressure driven 
 gas outflux being compensated by the gas influx
at the electrode.

In summary, we have suggested a dynamic equilibrium 
stabilization mechanism
for surface nanobubbles: The gas outflux driven by Laplace pressure
is compensated by a gas influx at the contact line, leading to
a metastable equilibrium under certain conditions.
A necessary ingredient into this model is a contact angle decrease
with decreasing bubble size $R$. Though such a decrease has been
observed both experimetally \cite{kam08a,kam08b} and numerically
\cite{che03,che06}, further work is required to better 
quantify and understand this dependence $\theta(R)$.

\noindent
{\it Acknowledgements:} 
We thank the participants of the Leiden Lorentz Center Workshop
on the Physics of Micro- and Nanofluidics (June 2008) for very helpful discussions. 
Special thanks goes to Bram Borkent for continuous stimulating discussions on
nanobubbles. 
MPB acknowledges support from the NSF Division of Mathematical Sciences
and the Harvard MRSEC grant.


\end{document}